\documentclass[fleqn,10pt]{wlscirep}
\usepackage[utf8]{inputenc}
\usepackage[T1]{fontenc}
\usepackage{graphicx}
\usepackage{amssymb}
\usepackage{amsfonts}
\usepackage{bbm}
\usepackage{amsmath}
\usepackage{lmodern}
\usepackage{mathrsfs}
\usepackage[framed,numbered,autolinebreaks,useliterate]{mcode}

\newcommand{\idol}{\ensuremath{\mathbbm 1}}

\newcommand{\tr}{{\rm Tr}}

\title{Numerical and analytical results for geometric measure of coherence and geometric measure of entanglement}

\author[1]{Zhou Zhang}
\author[1]{Yue Dai}
\author[1,*]{Yu-Li Dong}
\author[2,1,*]{Chengjie Zhang}
\affil[1]{School of Physical Science and Technology, Soochow University, Suzhou, 215006, China}
\affil[2]{School of Physical Science and Technology, Ningbo University, Ningbo, 315211, China}

\affil[*]{yldong@suda.edu.cn}

\affil[*]{chengjie.zhang@gmail.com}

\begin{abstract}
Quantifying coherence and entanglement is extremely important in quantum information processing. Here, we present numerical and analytical results for the geometric measure of coherence, and also present numerical results for the geometric measure of entanglement. On the one hand, we first provide a semidefinite algorithm to numerically calculate geometric measure of coherence for arbitrary finite-dimensional mixed states. Based on this semidefinite algorithm, we test randomly generated single-qubit states, single-qutrit states, and a special kind of $d$-dimensional mixed states. Moreover, we also obtain an analytical solution of geometric measure of coherence for a special kind of mixed states. On the other hand, another algorithm is proposed to calculate the geometric measure of entanglement for arbitrary two-qubit and qubit-qutrit states, and some special kinds of higher dimensional mixed states. For other states, the algorithm can get a lower bound of the geometric measure of entanglement. Randomly generated two-qubit states, the isotropic states and the Werner states are tested. Furthermore, we compare our numerical results with some analytical results, which coincide with each other.
\end{abstract}
\begin{document}

\flushbottom
\maketitle
%
%
\thispagestyle{empty}


\section*{Introduction}

Quantum coherence and entanglement are two basic concepts in quantum information theory, which are extensively applied to quantum information processing and quantum computational tasks \cite{g1}. Moreover, both quantum coherence and entanglement can be regarded as quantum resources, and they are useful for quantum-enhanced metrology, quantum key distribution and so on \cite{g27,g28,g29,g30,g31,g34,g35}.
Therefore, characterizing and quantifying coherence and entanglement become significant parts in quantum information theory \cite{g7}.

Quantum coherence is defined for a single system, and is widely used in quantum optics in previous studies \cite{h1,h2,h3,h4,h5,h6,h7,h8}. For any distance measure $D$ between two arbitrary quantum states, a general coherence measure is defined as $C_{D}(\rho)=\min_{\delta\in\mathbb{I}}D(\rho,\delta)$, i.e., the minimum distance from $\rho$ to all possible incoherent states $\delta\in\mathbb{I}$, where $\mathbb{I}$ is the set of all incoherent states \cite{g13,g14,g15,g38,j4}. From this definition, one can see that $C_{D}(\rho)=0$ if and only if $\rho$ is an incoherent state. Many distance-based coherence measures are proposed such as geometric measure of coherence, relative entropy of coherence and $l_{p}$ norm of coherence. The geometric measure of coherence is defined by using the fidelity between the measured state $\rho$ and its nearest incoherent state \cite{g3}. The relative entropy of coherence is another distance-based coherence measure \cite{g13,j1,g32}. Considering the coherence measures based on the matrix norms, the $l_{1}$ norm of coherence was introduced and studied in Ref. \cite{g13}. Besides, other different coherence measures have also been proposed \cite{u1,g32,u3,Da}. Furthermore, many experimental results on coherence have been reported \cite{exp1,exp2,exp3,exp4,exp5}.

Quantum entanglement is widely regarded as an essential feature of quantum mechanics, and entanglement measures have many applications \cite{app1,app2,app3,app4}. A class of entanglement measures are based on the fact that the closer a state is to the set $\mathbb{S}$ of separable states, the less entanglement it has \cite{g14,j4}. According to the distance measure $D$ between quantum states $\rho$ and $\sigma$, it is defined as $E_{D}(\rho)=\min_{\sigma\in\mathbb{S}}D(\rho,\sigma)$, i.e., the measure is the minimum distance to all possible separable states
\cite{g13,g14,g15,g38,j4}. One fundamental distance-based entanglement measure is the relative entropy of entanglement \cite{j4}, which can be considered as a strong upper bound for entanglement of distillation \cite{ra}. Another one is the geometric measure of entanglement (GME) \cite{GME1,GME2,wei}. Furthermore, the expected value of
entanglement witnesses can be used to estimate the GME \cite{og,zhang1,zhang2}. Other different entanglement measures have been proposed for multipartite systems and mixed states \cite{g14,g15}.

Most quantum states have no corresponding analytical solutions for the coherence and entanglement measures, so numerical algorithms must be applied. In some entanglement measures, several numerical algorithms have been used to solve related problems \cite{wa,wa1,wa2,g36}. Moreover, computing many entanglement
measures is NP hard for a general state \cite{sg,yh}, so some upper and lower bounds are proposed to describe entanglement \cite{k2,k3,k4,k5,k6,k7,k8,k9} and coherence measures \cite{pi,cn,g4,sap,hu}. In Refs. \cite{g2,see}, a semidefinite program (SDP) was proposed to calculate the fidelity between two states, so the geometric measures of coherence and entanglement probably can be numerically obtained based on the semidefinite program. In the following, we will try to provide semidefinite programs, in order to get the numerical results of the geometric measures of coherence and entanglement.

In this work, we first review the definition and properties of fidelity and its semidefinite program. Then we present numerical and analytical results for the geometric measure of coherence. Our algorithm can be used to numerically obtain the geometric measure of coherence for arbitrary finite-dimensional states. We test our semidefinite program for single-qubit states,  single-qutrit states, and a special kind of $d$-dimensional mixed states. For the special kind of $d$-dimensional mixed states, we also obtain an analytical solution of its geometric measure of coherence. Furthermore, we also propose another algorithm for the geometric measure of entanglement, which can obtain the geometric measure of entanglement for arbitrary two-qubit and qubit-qutrit states, and some special kinds of higher dimensional mixed states. For other states, the algorithm can get a lower bound of the geometric measure of entanglement.

\section*{Results}

\textbf{A semidefinite program for computing fidelity.}
We first review the fidelity and its semidefinite program in the following. The fidelity between states $\rho$  and $\chi$ is defined as \cite{g1}
\begin{equation}\label{f1}
F(\rho,\chi)\equiv\tr{\sqrt{\rho^{\frac12} \chi \rho^{\frac12}}}.
\end{equation}
For a pure state $|\psi\rangle$ and an arbitrary state $\chi$, one can get that
\begin{equation}\label{f2}
 \begin{split}
 F(|\psi\rangle,\chi)=\sqrt{\langle\psi|\chi|\psi\rangle}.
  \end{split}
\end{equation}
In Refs. \cite{g2,see}, Watrous and colleagues proposed a semidefinite program, whose optimal value equals the fidelity for given positive semidefinite operators, i.e., considering the following optimization problem
\begin{align}\label{f3}
 \mathrm{maximize}: \quad & \frac12 \tr(X)+ \frac12 \tr(X^\dagger)\notag ,\\  \mathrm{subject \  to} :\quad & \left( \begin{matrix}
    \rho &   X \\
    X^\dagger &  \chi \\
 \end{matrix}\right)\geq 0, \\
   \notag & X\in L(\mathbb{X}),\\
   \notag & \rho,\chi\in Pos(\mathbb{X}),
  \end{align}
where $L$ is the collection of all linear mappings. In a $\mathbb{X}$ complex Hilbert space, $Pos(\mathbb{X})$ is the set of positive semidefinite operators operating on $\mathbb{X}$. Then the maximum value of $\frac12 \tr(X)+ \frac12 \tr(X^\dagger)$ is equal to $F(\rho,\chi)$. $X$ is a randomly generated complex matrix of the same order as $\chi$.

The SDP can not only solve the problem effectively, but also prove the global optimality under weak conditions \cite{LS}. This implies that SDP optimization problems can be tackled with standard numerical packages. In this paper, the optimization of the SDP (\ref{f3}) can be solved by using the Matlab parser YALMIP \cite{g19} with the solvers, SEDUMI \cite{g20} or SDPT3 \cite{g21,trh}. In fact, there exist serval SDP problems in quantum information theory. For example, SDP programs have been used in entanglement detection and quantification \cite{Doherty1,Doherty2,Doherty3,Toth,Jungnitsch,g12}, quantifying quantum resources \cite{Roope}. Furthermore, the SDP (\ref{f3}) has also been used for calculating the fidelity of quantum channels \cite{yuan}.

\vspace{8pt}
\textbf{Geometric measures of coherence.}
In a $d$-dimension Hilbert space $\mathcal{H}$ with its corresponding reference basis $\{|i\rangle\}_{i = 0}^{d-1}$, a state is incoherent if and only if it is a diagonal density matrix under the reference basis \cite{g27,g28}. All incoherent states can be represented as \cite{g27,g28}
\begin{equation}\label{delta1}
\delta = \sum_{i=0}^{d-1}   p_i|i\rangle\langle i|.
\end{equation}
Thus, the geometric measure of coherence is defined as \cite{g3}
\begin{equation}\label{f4}
C_{g}(\rho)=1-[\max_{\delta\in\mathbb{I}}F(\rho,\delta)]^2,
\end{equation}
with the maximum being taken over all possible incoherent states $\delta\in\mathbb{I}$. Based on the SDP (\ref{f3}), Eqs. (\ref{delta1}) and (\ref{f4}), we provide the MATLAB
code for the semidefinite program of geometric measure of coherence in Supplemental Material.

For an arbitrary single-qubit state $\rho$, its analytical solutions of $C_{g}(\rho)$ has been derived \cite{g3}
\begin{equation}\label{f5}
C_{g}(\rho)=\frac12(1-\sqrt{1-4|\rho_{01}|^2}),
\end{equation}
where $\rho_{01}$ is the off-diagonal element of $\rho$ in reference basis. To compare with this analytical solution, we randomly generate $10^5$ density matrices and calculate their $C_{g}(\rho)$ by analytical and numerical methods, respectively. The analytical results are calculated based on Eq. (\ref{f5}), and the numerical results are obtained by optimizing the semidefinite program \cite{SM}. The maximum deviation between the analytical and numerical results is $3.19 \times 10^{-9}$.

For a pure state  $|\psi\rangle=\sum_i \lambda_i |i\rangle$, its geometric measure of coherence is
\begin{equation}\label{f6}
C_{g}(|\psi\rangle)=1-\max_{i}\{|\lambda_i|^2\},
\end{equation}
where $|\lambda_i|^2$ is the diagonal elements of $|\psi\rangle\langle\psi|$ \cite{g28}. However, the corresponding analytical solutions of $C_{g}(\rho)$ are difficult to calculate for general mixed states, so it is necessary to get some lower and upper bounds of $C_{g}(\rho)$. Here we employ the lower and upper bounds proposed in Ref. \cite{g4} and compare them with our numerical results of the optimization program. For a general $d\times d$ density matrix $\rho$, its $C_{g}(\rho)$ satisfies \cite{g4}
\begin{equation}\label{f7}
1-\frac{1}{d}-\frac{d-1}{d}\sqrt{1-\frac{d}{d-1}(\tr{\rho^2}-\sum_{i}{\rho_{ii}^2})} \leq C_{g}(\rho) \leq \min \{1-\max_{i}\{\rho_{ii}\},1-\sum_{i}{b_{ii}^2} \},
\end{equation}
where $b_{ii}$ is from $\sqrt{\rho}=\sum_{ij}b_{ij}{|i\rangle\langle j|}$.

\begin{figure}
\includegraphics[scale=0.4]{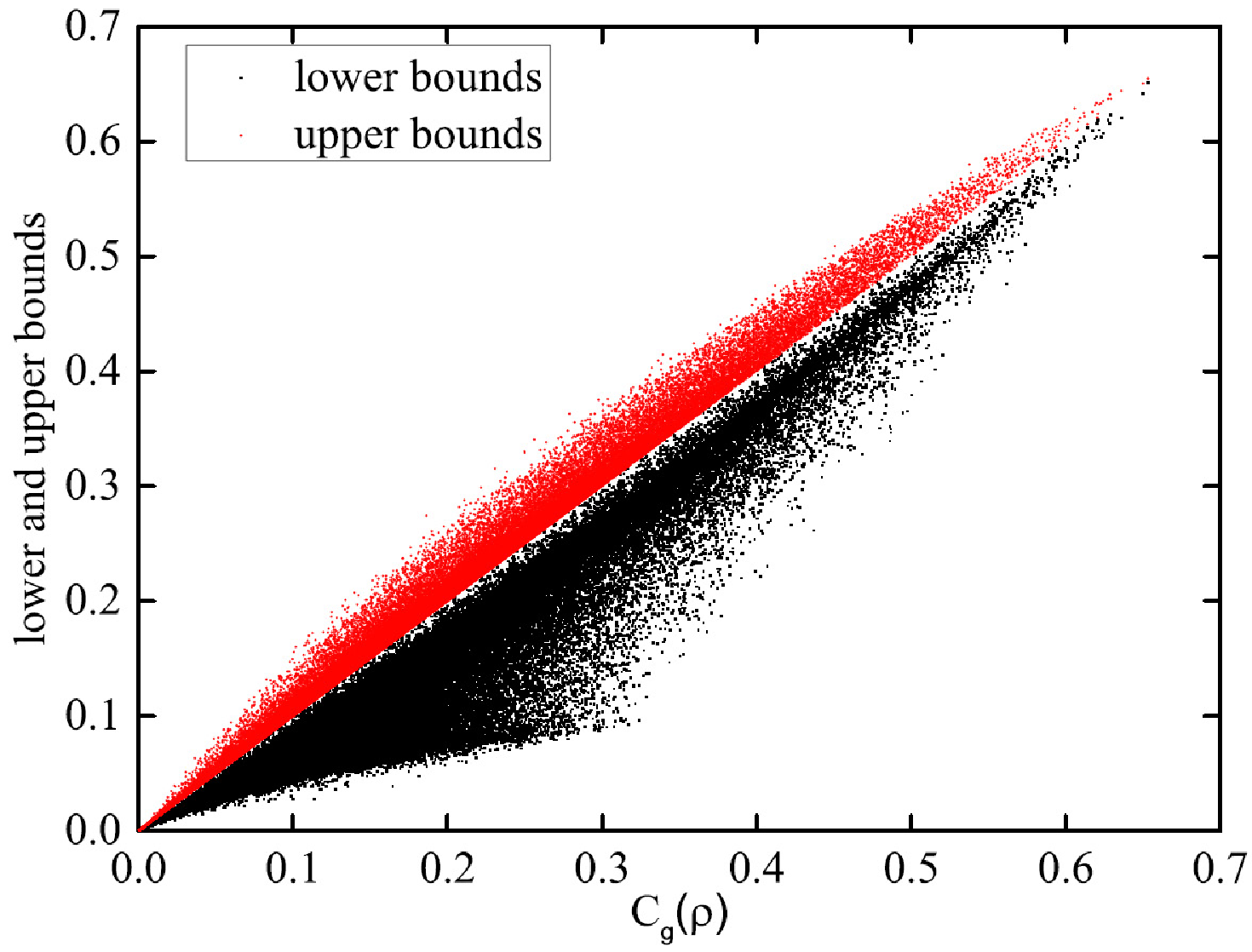}
\caption{Black points (Red points) represent the left (right) hand side of the inequality (\ref{f7}). $C_{g}(\rho)$ indicates our numerical results. There is apparent dividing line between them.}\label{fig1}
\end{figure}

Since there is no corresponding analytic solution for general single-qutrit mixed states, we randomly generate $10^5$ density matrices to draw their corresponding upper and lower bounds. In Fig. \ref{fig1}, there is a clear dividing line between two bounds indicating that the numerical results obtained by our algorithm coincide with the analytical results from the inequality (\ref{f7}), and points on the upper bound is closer to the dividing line than points on the lower bound for many  $3\times 3$ density matrices.

Now we consider the following mixed state
\begin{equation}\label{f31}
\rho=p |\psi^+\rangle\langle\psi^+|+(1-p)\frac{\idol}{d},
\end{equation}
with $|\psi^+\rangle=\frac{1}{\sqrt{d}}\sum_{i=0}^{d-1}|i\rangle$, $\idol$ being the $d\times d$ identity matrix, and $0\leq p \leq 1$. Since the mixed state $\rho$ is highly symmetric, it will remain unchanged when we exchange its basis order. It limits that the reference incoherent state in geometric measure of coherence must have the same diagonal elements, i.e., the closest incoherent state $\delta$ to the density matrix $\rho$ has to be
\begin{equation}\label{f30}
\delta=\sum_{i=0}^{d-1}\frac{1}{d}|i\rangle\langle i|.
\end{equation}
Therefore, we can obtain its analytical solution of geometric measure of coherence.

\textit{Proposition 1.} For the mixed state $\rho=p |\psi^+\rangle\langle\psi^+|+(1-p)\frac{\idol}{d}$ with  $|\psi^+\rangle=\frac{1}{\sqrt{d}}\sum_{i=0}^{d-1}|i\rangle$, its analytical solution of geometric measure of coherence is
\begin{equation}\label{f34}
C_{g}(\rho)=1-\frac{1}{d^2}[(d-1)\sqrt{1-p}+\sqrt{1+(d-1)p}]^2.
\end{equation}

\begin{figure}
\includegraphics[scale=0.6]{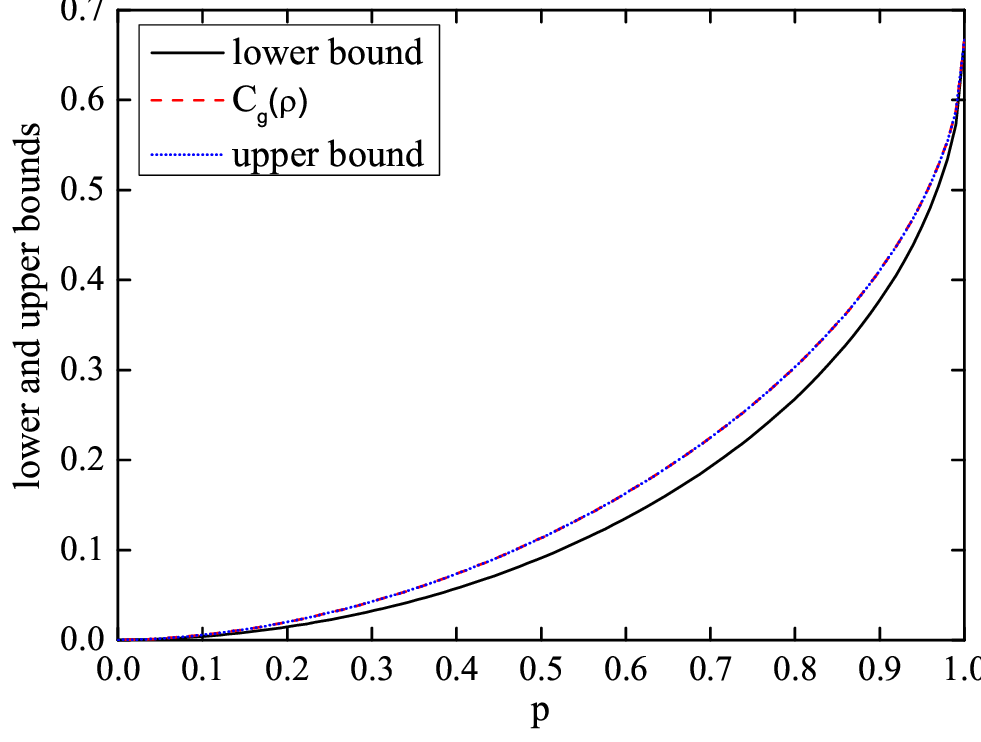}
\caption{The black line (blue dots) indicates the left (right) hand side of the inequality (\ref{f7}). The red dashed line $C_{g}(\rho)$ from Eq. (\ref{f34}) is coincident with the upper bound for $d=3$.}\label{fig2}
\end{figure}

\begin{figure}
\includegraphics[scale=0.6]{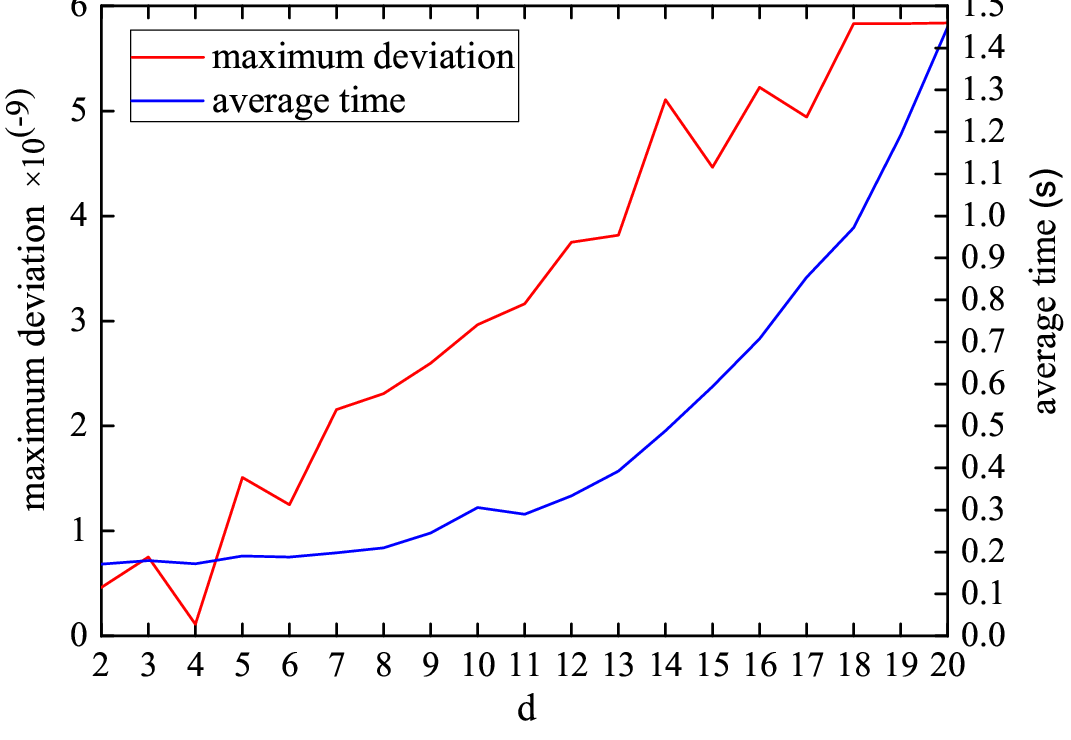}
\caption{The red line represents the maximum deviation between the numerical solution and the analytical solution. The blue line indicates the average time $\overset{\sim}{t} (s)$ of each operation for the density matrices
(\ref{f31}).}\label{fig3}
\end{figure}

This analytical result is equal to its corresponding upper bound which is the right hand side of the inequality (\ref{f7}). When $2\leq d \leq 20$, we calculate their analytical and numerical results as well as maximum deviation between them. For $d=3$ the corresponding graph is drawn and the rest have the similar phenomena like it. In Fig. \ref{fig2}, $C_{g}(\rho)$ and its upper bound coincide for $d=3$, and the maximum deviation between them is $1.51\times 10^{-9}$. In Fig. \ref{fig3}, the maximum deviation between the numerical and analytical results is about $10^{-9}$ orders of magnitude. Although the average time $\overset{\sim}{t} (s)$ of each operation increases exponentially, it is within an acceptable range in the low dimensional case.

\vspace{8pt}
\textbf{Geometric measures of entanglement.}
A separable bipartite pure state can be written in the following product form
\begin{equation}\label{1abel5}
|\psi^{AB} \rangle=|\psi^{A} \rangle \otimes |\psi^{B} \rangle.
\end{equation}
For mixed states, if it can be represented as convex weights $p_{i}$ and product states $\rho_{i}^{A}\otimes \rho_{i}^{B}$ \cite{we}
\begin{equation}\label{label1}
\rho^{AB}=\sum_{i}p_{i}\rho_{i}^{A}\otimes \rho_{i}^{B},
\end{equation}
then $\rho^{AB}$ is called separable.
For a bipartite state, if there is no negative eigenvalues after the partial transposition of subsystem $A$, this bipartite state is called the PPT states \cite{g6}, i.e., a bipartite state
\begin{equation}\label{f15}
\rho_{A|B}=\sum_{ij,kl} \rho_{ij,kl} |i\rangle\langle j|_{A} \otimes |k\rangle\langle l|_{B},
\end{equation}
is PPT, when its partial transposition with respect to the subsystem $A$ satisfys
\begin{equation}\label{f16}
 \rho_{A|B}^{T_{A}}=\sum_{ij,kl} \rho_{ij,kl}|j\rangle\langle i|_{A} \otimes |k \rangle\langle l|_{B}  \geq 0.
\end{equation}

The GME is defined as follows  \cite{AS}
\begin{equation}\label{f8}
E_{G}(\rho)=1-[\max_{\sigma\in\mathbb{S}}F(\rho,\sigma)]^2,
\end{equation}
where $\mathbb{S}$ is the set of all separable states. We replace $\mathbb{S}$ with the set $\mathbb{P}$ of all PPT states, because $\mathbb{S}$ cannot be easily expressed in the semidefinite programs, but $\mathbb{P}$ can be expressed since for a given density matrix one can directly calculate its partial transpose \cite{og}. Thus, based on the fact that $\mathbb{S}$ is a subset of $\mathbb{P}$ \cite{g6}, one can obtain a lower bound of $E_{G}(\rho)$, i.e.,
\begin{equation} \label{f10}
E_{G}(\rho)\geq \overset{\sim}{E}_{G}(\rho),
\end{equation}
where the lower bound $\overset{\sim}{E}_{G}(\rho)$ is defined by
\begin{equation}\label{f9}
\overset{\sim}{E}_{G}(\rho)=1-[\max_{\tilde{\sigma}\in\mathbb{P}}F(\rho,\tilde{\sigma})]^2.
\end{equation}
The equality in Eq. (\ref{f10}) holds for all two-qubit and qubit-qutrit states  \cite{hm}, and some special kinds of higher dimensional mixed states.
Based on the SDP (\ref{f3}), Eqs. (\ref{f16}) and (\ref{f8}), we provide the MATLAB
code for the semidefinite program of $\overset{\sim}{E}_{G}(\rho)$ in Supplemental Material.

For pure states, the GME is defined as \cite{g7}
\begin{equation}\label{f11}
E_{G}(|\psi\rangle)=1-\max_{|\phi\rangle \in\mathbb{S}}|\langle\psi|\phi\rangle|^2,
\end{equation}
Moreover, it is defined via the convex roof construction for mixed states. If $\rho$ is a two-qubit state, the corresponding expression of $E_{G}(\rho)$ is \cite {g7,g8,g9}
\begin{equation}\label{f12}
E_{G}(\rho)=\frac12 (1-\sqrt{1-C(\rho)^2}).
\end{equation}
The $C(\rho)$ is called concurrence that its expression is
\begin{equation}\label{f13}
C(\rho)=\max{\{0,\lambda_{1}-\lambda_{2}-\lambda_{3}-\lambda_{4}\}},
\end{equation}
where $\{\lambda_{i}\}$ are the square root of eigenvalues of $\rho\overset{\sim}{\rho}$ in descending order and $\overset{\sim}{\rho}=(\sigma_{y}\otimes\sigma_{y})\rho^\ast(\sigma_{y}\otimes \sigma_{y})$. In order to compare the analytical result Eq. (\ref{f12}) and the numerical result for two-qubit states, we randomly generate $10^5$ density matrices and calculate the analytical and numerical results respectively. The maximum difference between them is $1.57 \times 10^{-9}$.
\begin{table}
\begin{tabular}{c|cccc}
\hline
\hline
d & 2 & 3 & 4 & 5  \\
  \hline
  $\Delta E_{G}(\rho)$ & $1.03 \times 10^{-9}$ & $2.02 \times 10^{-9}$ & $3.21 \times 10^{-9}$ & $5.05\times 10^{-9}$ \\
  \hline
  $\overset{\sim}{t} (s)$ & $0.29$ & $0.49$ & $2.23$ & $27.35$ \\
  \hline
  \hline
\end{tabular}
 \caption{The maximum deviation between the numerical solution and the analytical solution with $\Delta E_{G}(\rho)=E_{G}(\rho)- \overset{\sim}{E}_{G}(\rho)$, and the average time $\overset{\sim}{t} (s)$ of each operation for the isotropic states (\ref{f14}).}\label{lab2}
\end{table}

Now we apply our semidefinite program to the isotropic states, where the forms of these states are \cite{g7}
\begin{equation}\label{f14}
\rho=\frac{1-F}{d^2-1}(\idol-|\Phi^{+}\rangle\langle \Phi^{+}|)+F|\Phi^{+}\rangle\langle \Phi^{+}|,
\end{equation}
with the maximally entangled state $|\Phi^{+}\rangle=\frac{1}{\sqrt{d}}\sum_{i=0}^{d-1}|ii\rangle$ and $0\leq F\leq 1$. The analytical solutions for the GME of these isotropic states were given in Ref. \cite{g7}, and the states are separable if and only if $F \leq \frac{1}{d}$ \cite{g10}. For Eq. (\ref{f14}) when $2 \leq d \leq 5$, we calculate that the maximum deviation between the numerical and analytical solution by our semidefinite program and the analytical solution given in \cite{g7}, respectively. The results are summarized in Table \ref{lab2}, where $\overset{\sim}{t} (s)$ denotes the average time of each operation. In the example tested above, the semidefinite program always obtain the same value as $E_{G}(\rho)$ within the precision given in Table \ref{lab2}.

\begin{table}
\begin{tabular}{c|cccc}
\hline
\hline
d & 2 & 3 & 4 & 5  \\
  \hline
  $\Delta E_{G}(\rho)$ & $5.00 \times 10^{-10}$ & $2.26 \times 10^{-9}$ & $3.17 \times 10^{-9}$ & $5.57\times 10^{-9}$ \\
  \hline
  $\overset{\sim}{t} (s)$ & $0.32$ & $0.46$ & $1.47$ & $14.83$ \\
  \hline
  \hline
\end{tabular}
 \caption{The maximum deviation between the numerical solution and the analytical solution with $\Delta E_{G}(\rho)=E_{G}(\rho)- \overset{\sim}{E}_{G}(\rho)$, and the average time $\overset{\sim}{t} (s)$ of each operation for the Werner states (\ref{f32}).}\label{lab3}
\end{table}

Finally, we apply semidefinite program to the Werner states that it can be expressed as a linear combination of two operators of the \textit{identity} $\idol$ and the \textit{swap} $\hat{F}\equiv \sum_{ij} |ij\rangle\langle ji|$\cite{g7}, i.e., $\rho=a \idol+b \hat{F}$, where $a$ and $b$ are both real coefficients and are limited by $\tr{\rho}=1$. When one of the parameters is considered, the states can be expressed as
\begin{equation}\label{f32}
\rho=\frac{d^2-fd}{d^4-d^2}\idol\otimes \idol+\frac{fd^2-d}{d^4-d^2} \hat{F},
\end{equation}
with $f\equiv \tr(\rho \hat{F})$. The corresponding analytic solution for the Werner states (\ref{f32}) is \cite{g7}
\begin{equation}\label{f33}
E_{G}(\rho)=\frac{1}{2}(1-\sqrt{1-f^2}),
\end{equation}
where $f \leq 0$ and $0$ otherwise. For $2 \leq d \leq 5$, we apply our semidefinite program to states (\ref{f32}), so the maximum deviation between the numerical solution and the analytical solution is calculated, respectively. The results are summarized in Table \ref{lab3}, where $\overset{\sim}{t} (s)$ is the average time of each operation.

\section*{Discussion}
For the geometric measure of entanglement, we compare our algorithm with the algorithm proposed in Ref. \cite{g36}. Streltsov and colleagues proposed an algorithm \cite{g36}, which can be easily implemented by solving an eigenproblem or finding a singular value decomposition of a matrix. However, their algorithm needs the iteration with many steps, and it may converge to a local minimum which is not the exact value of the geometric measure of entanglement. Our algorithm, which does not need iteration and has no local minimum problems, is based on  semidefinite program and easy to implement. Unfortunately, the shortcoming of our algorithm is also obvious. It can be used to calculate the geometric measure of entanglement for arbitrary two-qubit and qubit-qutrit states, and some special kinds of higher dimensional mixed states. But for other states, our algorithm can only get a lower bound of the geometric measure of entanglement.

In this paper, we introduced numerical and analytical results to compute the geometric measures of coherence and the entanglement. In coherence measures, the deviation between the numerical solution and the analytical solution was an order of magnitude of $10^{-9}$ for single-qubit states. Furthermore, we obtained the analytical solution of the geometric measure of coherence  $C_{g}(\rho)=1-\frac{1}{d^2}[(d-1)\sqrt{1-p}+\sqrt{1+(d-1)p}]^2$ for the special kind of mixed states $\rho=p |\psi^+\rangle\langle\psi^+|+(1-p)\frac{\idol}{d}$. For randomly generated $3$-dimensional density matrices, we have drawn a boundary diagram with a apparently clear boundary line. In entanglement measures, we used PPT states to replace the set of separable states and calculated two-qubit states, the isotropic states and the Werner states by using fidelity and its semidefinite program, and then concluded that their maximum deviation is almost on the order of magnitude of $10^{-9}$.





\section*{Acknowledgements}

C.Z. gratefully acknowledges Otfried G\"uhne and Haidong Yuan for helpful discussions. This work is supported by the National Natural Science Foundation of China (Grant No. 11734015), and K.C.Wong Magna Fund in Ningbo University.

\section*{Author contributions statement}

Z.Z., Y.-L.D. and C.Z. wrote the main manuscript text and Y.D. prepared figures 1-3. All authors reviewed the manuscript.

\section*{Additional information}

Competing financial interests: The authors declare no competing interests.

\section*{Corresponding authors}
Correspondence and requests for materials should be addressed to Y.-L.D. (email:yldong@suda.edu.cn) or C.Z. (email:chengjie.zhang@gmail.com).



\section*{Supplemental Material}

Here we provide some detailed calculations of the single-qubit states, randomly generate $3$-dimensional density matrices, and a special kind of $d$-dimensional density matrices for coherence measures. The corresponding MATLAB code for the semidefinite program of geometric measure of coherence is
\begin{lstlisting}
function [C, D]=coherence(rho)

   d=length(rho);

   delt=sdpvar(d,d, 'diagonal', 'real');

   X=sdpvar(d,d, 'full', 'complex');

   constr=[[rho X; ctranspose(X) delt]>=0, delt>=0, trace(delt)==1];

   result=solvesdp(constr, -trace(X)-trace(ctranspose(X)), sdpsettings('verbose', 1));

%check for errors
   if (result. problem ~= 0 )
      disp(result. info);
   end

%return the maximum fidelity
   F=double((trace(X)+trace(ctranspose(X)))/2);

   D=double(delt);

   C=1-F^2;

end
\end{lstlisting}
We have used the parser YALMIP \cite{syalmip} with the solvers, SEDUMI \cite{ssedumi} or SDPT3 \cite{ssdpt3}.

We also provide some detailed calculations of the two-qubit states, the isotropic states and the Werner states for entanglement measures. The corresponding MATLAB code for the semidefinite program of geometric measure of entanglement is
\begin{lstlisting}
function [E, E2]=GME(rho)

%d(i) is the dimension of the matrix
   d1=2;
   d2=2;

   sigma=sdpvar(d1*d2, d1*d2, 'hermitian', 'complex');

   X=sdpvar(d1*d2, d1*d2, 'full', 'complex');

%performing partial transpose for sigma and the pt subroutine comes from pptmixer.
   constr=[[rho X; ctranspose(X) sigma]>=0, sigma>=0, pt(sigma, [1,0], [d1, d2])>=0, trace(sigma)==1];

   result=solvesdp(constr, -trace(X)-trace(ctranspose(X)), sdpsettings('verbose', 1));

%check for errors
   if (result. problem ~= 0)
	   disp(result. info);
   end

%return the maximum fidelity
   F=double((trace(X)+trace(ctranspose(X)))/2);

   D=double(sigma);

   E=1-F^2;

end
\end{lstlisting}
We have used the pt subroutine from the program PPTmixer, where the PPTmixer was presented in \cite{pt} (or \cite{ppt}).

\end{document}